\documentstyle [12pt]{article}
\begin{document}
\baselineskip = 21 pt
 \thispagestyle{empty}
 \title{
\vspace*{-2.5cm}
\begin{flushright}
\begin{tabular}{c c}
& {\normalsize MPI-Ph/93-10}\\
& {\normalsize February 1993}
\end{tabular}
\end{flushright}
\vspace{1.5cm}
 On the Unification of Couplings in the Minimal
Supersymmetric Standard Model \\
 ~\\}
 \author{M. Carena, S. Pokorski\thanks{On leave of absence
         from the Univ. of Warsaw.} $\;$
         and  C. E. M. Wagner\\
 ~\\
Max-Planck-Institut f\"{u}r Physik\\
Werner-Heisenberg-Institut\\
F\"{o}hringer Ring 6\\
D-8000 M\"{u}nchen 40, Germany.\\
 ~\\ }
\date{
\begin{abstract}
The unification of gauge and Yukawa couplings within the
minimal supersymmetric standard model is studied at the two
loop level. We derive an expression  for the effective  scale,
$T_{SUSY}$, which characterizes the supersymmetric particle
threshold corrections to the gauge couplings, and demonstrate
that $T_{SUSY}$  is only slightly dependent on the   squark
and slepton masses, and  strongly dependent on the Higgsino
masses as well as on the mass ratio of the gauginos of the strong
and weak interactions. Thus, if the gaugino masses proceed from
a common soft  supersymmetry breaking term at the unification scale,
and there is no significant gaugino-Higgsino mixing, $T_{SUSY}$ will
have a strong  dependence on the supersymmetric mass parameter $\mu$.
Moreover, the value of the top quark Yukawa coupling necessary to
achieve the unification of bottom and tau Yukawa couplings is also
governed by $T_{SUSY}$, and it yields predictions for the top quark
mass which are close to the quasi infrared fixed point results
associated with the triviality bounds on this quantity.
{}From the requirement of perturbative consistency of the
top quark Yukawa sector of the theory, we obtain constraints
on the allowed splitting of the supersymmetric spectrum, which,
for certain values of the running bottom quark mass, $m_b$,
are stronger than those ones coming from the experimental
constraints on the strong gauge coupling.  For example, for
$m_b(M_b) \leq 4.1$ GeV, which approximately corresponds to
a physical mass $M_b \leq 4.7$ GeV, we obtain that,
 for moderate values of $\tan\beta$, perturbative
unification may be achieved only if $\alpha_3(M_Z) \leq 0.124$.
\end{abstract}}
\maketitle

\newpage
\section{Introduction}

\hspace*{0.6cm}
The Standard Model (SM) provides a very good understanding of the
strong and electroweak interactions and, so far, it has withstood all
the experimental onslaugths.
Yet, there are a host of open questions, which need to be answered.
Some of them are of technical nature, since they are related to the
probation of the model at a higher level of refinement upon
more precise measurements from the experiments and
more meticulous
 theoretical computations to analyze the discrepancies between
observations and predictions.  Apart from this type of questions,
there
are others which involve more fundamental issues,
 and are those concerning
the generation of masses with the associated large number of unknown
parameters,  the properties of matter with the unexplained apparent
triplication of fermions into families and the peculiar assignment
of quantum numbers and, most challenging, the mystery of forces.
In fact, the SM does not comprise all the forces observed in nature.
Moreover, even if there were
no other forces in nature, let aside gravity,
besides those associated with SU(3)$_c$ $\times$ SU(2)$_L$ $\times$ U(1),
it would be
natural to ask about the physical principle which selects
this group structure  above others.
As a consequence,
the idea of the existence of an underlying structure unifying all the
gauge interactions,  from which the SM is derived as an effective
theory, appears as a sensible alternative.
In fact, using the Renormalization Group (RG) evolution
to determine the unknown high energy behavior of
the SM parameters,
it is possible to envisage that the three
different gauge couplings may converge to
 a common value at some unexplored
high energy scale $\Lambda$ \cite{earlySM},\cite{early}.
 Indeed, the unification of couplings is the
 greatest conceptual advantage of
 a Grand Unified Theory (GUT). Other quite important consequence of a
GUT is that it naturally implies charge quantization.

The efficiency of the RG approach to explore   the short
distance physics is, of course, strongly
dependent on the accuracy of the low energy data
to be used as the initial boundary conditions in the RG evolution.
In this respect,
considering the available accurately determined value of the
electromagnetic gauge coupling as well as
 the precision measurements
for the weak couplings from LEP and from neutrino scattering experiments
and in spite of
the rather large experimental uncertainties in the determination of
the strong gauge coupling, it has been already shown
in the literature
that the simplest GUT theories, including SU(5), are ruled out.
On the contrary,
after the inclusion of supersymmetry,
the  gauge couplings of the SM  converge to a common value
at a high energy scale which may be  sufficiently high to prevent a
too fast proton decay rate,
incompatible with experiment \cite{ABF} - \cite{LNPZ}.

As a matter of fact, recently there has been  renewed interest
in analyzing models of GUT's, and, indeed, some
possible high energy scenarios have been studied
\cite{AN} - \cite{DHR}.
However, most important is to observe
that all these various grand unification scenarios
lead to similar low energy predictions.
Thus, for the extent of this work, we shall concentrate on the minimal
supersymmetric extension of the SM in the light of the minimal
presumption of Grand Unification, without any assumptions about
the physics above the unification scale.
 In particular, we shall also analyze
 the unification of couplings
in the Yukawa sector, which appears naturally in some grand
unified models \cite{Ramond}-\cite{DHR}.
The equality of the bottom quark and $\tau$ lepton Yukawa couplings
is obtained in models with
SU(5) gauge symmetry, while the extreme case of equality of the
$\tau$ lepton  and the
top and bottom quark  Yukawa couplings may occur in
SO(10) and E(6) unification schemes.
Beyond all the details
depending on  a particular high energy model,  which we shall not address
in the present study, what remains effective is that definite predictions
for the top quark mass can be made in the framework of gauge and
Yukawa coupling  unification.

The issue of the unification of couplings in the Minimal
Supsersymmetric Standard Model (MSSM) has been
addressed to various extensions in previous analyses
\cite{ABF}-\cite{DHR}.
However, in the
present work, we study some aspects of the
Grand Unification framework, which yield
new interesting results in relation to the supersymmetric particle
spectrum and the
 supersymmetric threshold
usually considered in the literature.
   Our paper is organized as follows. In section 2 we present the
experimental predictions, which we shall consider to constrain
the values of the low energy parameters  of the theory.
 In  section 3, we describe
the two loop  RG method that
we  use in this study. In particular, we explain in detail
the treatment of the different threshold corrections which
are incorporated in our analysis. Most important, we derive
an expression for the effective
scale, $T_{SUSY}$, which governs the supersymmetric
threshold corrections as a function of the different
supersymmetric particle masses.
In section 4 we discuss the general features
of gauge and Yukawa coupling
unification. We present an approximate analytical
computation for the top quark mass derived from
the infrared quasi fixed point associated with the triviality
bounds on the top quark Yukawa coupling, and compare these
results with our numerical two loop predictions.
In section 5
we analyze the constraints on $T_{SUSY}$ and, therefore,
on the splitting of the
sparticle spectrum, which, depending on the exact values
of the bottom quark mass
and the weak mixing angle
$\sin^2 \theta_W$, may be derived  either from
bounds on the top quark Yukawa coupling coming from
the  requirement of perturbative consistency,
or from the experimental bounds on the strong gauge coupling,
$\alpha_3(M_Z)$.
Section 6 is devoted to the study of the two
loop RG results for the top quark mass as a function of
$\tan \beta$ - the ratio of the vacuum expectation values
of the two Higgs fields,
 and its dependence on $\alpha_3(M_Z)$
(or equivalently on $\sin^2 \theta_W$) and the bottom quark mass.
In particular, we analyze the conditions for which
the  three Yukawa couplings of the third generation acquire a common
 value at the unification scale,
 $M_{GUT}$.    We also present the predictions for the light Higgs mass
as a function of $\tan \beta$ and their
dependence on the squark mass.
We reserve section 7 for our conclusions.

\section{Experimental Constraints on the Low Energy Parameters}

   In order to analyze the unification of gauge couplings
in the MSSM,
and obtain reliable results from the RG analysis, it is necessary
to consider the experimental data, to determine
the low energy parameters as
precise as possible.
   Due to the most recent analysis of the LEP data,
the electroweak parameters are known at present with high
precision. The largest uncertainty comes from the
unknown top quark mass, which induces not only logarithmic
but also quadratic corrections on the value of
$\sin^2 \theta_W$. In the following, we
shall define the gauge couplings in
the modified minimal substraction
renormalization scheme ($\bar{MS}$), introduced
in Ref.\cite{DFS}. In such case,
the best fit to the data is obtained for \cite{Langacker}
\begin{equation}
\sin^2 \theta_W(M_Z)
 = 0.2324 \pm 0.0003,
\end{equation}
for a top quark mass value $M_t = 138$ GeV. If the top
quark mass is left free, the central value does not change,
but the errors increase. For other values of the top quark
mass,  the best fit to the data is obtained for values
of the weak mixing angle which are related to $M_t$ by
\begin{equation}
\sin^2 \theta_W(M_Z) \simeq
0.2324 - 1.\; 10^{-7} GeV^{-2}
 \left(M_t^2 - M_{t_0}^2\right),
\label{eq:sinmt}
\end{equation}
with $M_{t_0} = 138$ GeV. This implies that, for a top quark
mass $M_t \simeq 100$ GeV, the best fit is for
$\sin^2 \theta_W \simeq 0.2335$,
while, for $M_t \simeq 200$ GeV, one obtains
$\sin^2 \theta_W \simeq
0.2305$.
The value of the electromagnetic gauge coupling
is only logarithmically dependent on $M_t$ and for
$M_t = M_{t_0}$ is given by
\begin{equation}
\frac{1}{\alpha(M_Z)} = 127.9.
\end{equation}

 On the contrary
 to what happens in the electroweak sector of the
theory, the
strong gauge coupling $\alpha_3(M_Z)$ is not so
precisely known.
A general tendency in the experimental determination of
$\alpha_3(M_Z)$ is that those values which are obtained
from low energy experiments are slightly lower than those
obtained from LEP data. Although a light gluino could
improve the agreement between the low and the high
energy data \cite{Glu},
there are theoretical uncertainties
in both determinations, and
these uncertainties may account for the apparent discrepancies in the
value of $\alpha_3(M_Z)$ \cite{Altarelli}.
A conservative attitude is to take the value for the strong
gauge coupling
to be in the range \cite{Langacker}
\begin{equation}
\alpha_3(M_Z) = 0.12 \pm 0.01.
\end{equation}

The top quark has not yet been observed, but, as mentioned
above, within the SM, its mass may be indirectly
determined through its radiative correction effects, in
particular, through the measurement of the $\rho$ parameter.
Conservative upper and lower bounds on the physical
top quark mass  are given by
\begin{equation}
91 \; GeV \leq M_t \leq 200 \; GeV.
\end{equation}
The bottom quark, instead, has been already observed
experimentally. However, there is still a large
uncertainty about
its physical mass. Following the particle data
book \cite{Partd} we
shall take values for the physical bottom quark mass
in the range,
\begin{equation}
M_b = 4.7 - 5.2 \; GeV.
\end{equation}

 One should remember, that there is a significant
quantitative
difference between the running bottom (top) quark mass and the
corresponding
physical  mass, defined as the location of
the pole in its two point function. This difference comes
from  QCD  corrections, which, at the level of two loops,  are
given by,
\begin{equation}
m_q(M_q) = \frac{M_q}{1 + \frac{4 \alpha_3(M_q)}{3 \pi}
+ K_q \left( \frac{\alpha_3(M_q)}{\pi}\right)^2} ,
\label{eq:botmass}
\end{equation}
where
$M_q$ and $m_q$ are  the physical and
running  quark masses, with
 $q=b,t$ denoting  the bottom and top quarks,
respectively,
and $K_b=12.4$,
$K_t \simeq 11 $ \cite{RSM}.
{}From Eq.(\ref{eq:botmass}), we obtain
that, for $\alpha_3(M_Z) = 0.115$ $ (0.125)$ and a
physical bottom quark
mass $M_b = 4.7,4.9$ and 5.2 GeV, the corresponding running
mass parameter is
$m_b (M_b) \simeq 4.1 (4.0), 4.3 (4.2) $ and
4.6 (4.5) GeV, respectively. The bottom quark running
mass is typically a
15$\%$ lower
than the physical mass, with a slight dependence on
the exact value of the strong gauge coupling $\alpha_3(M_Z)$.
As we shall discuss below, assuming bottom and tau Yukawa coupling
unification,
the  difference between the
physical and running bottom quark
masses has relevant consequences on the
top quark
Yukawa sector. The reason is, that the bottom quark mass value
fixes the overall scale of the bottom quark
Yukawa coupling. Our predictions
are, instead, only slightly dependent on a moderate
 variation
of the energy scale at which the running mass is evaluated.

   In contrast to the third generation of quarks, the tau lepton
mass is very well known,
\begin{equation}
M_{\tau}
 = 1.78 \; GeV .
\end{equation}
In this case, the difference between the running and the
physical masses
 is only due to electromagnetic corrections
and can be neglected for the aim of the computations performed
in the present work.

\section{Renormalization Group Analysis and Threshold Corrections}

 A first approach to the problem of unification of couplings
within the context of the MSSM
is to assume the absence of threshold corrections at the grand
unification scale. In addition, a usual assumption considered in the
literature
is that all
the supersymmetric masses are fixed at a common energy scale, $M_{SUSY}$.
In such case, the
  effective scale  $T_{SUSY}$, which
  characterizes the supersymmetric threshold corrections to the gauge
couplings,
is directly equal to $M_{SUSY}$.
Once these assumptions are made,
for a given top quark mass, the value of the
weak mixing angle may be predicted as a function of the
strong gauge coupling $\alpha_3(M_Z)$ and the overall soft
supersymmetric threshold scale $M_{SUSY}= T_{SUSY}$.
Within the above
framework, it has been shown that, for values of the weak
mixing angle and the strong gauge coupling
consistent with experimental
data, the unification of couplings
may be achieved for values of the grand unification scale of
the order of $10^{16}$ GeV  and $M_{SUSY}$ of the order of
the weak scale \cite{ABF}.
Based on this fact, it has been first thought that a
more precise determination of the values of $\alpha_3(M_Z)$
and $\sin^2\theta_W(M_Z)$ would result in a prediction for
$T_{SUSY}$ and,  hence, for the characteristic value of the
supersymmetric particle mass $M_{SUSY}$.
 However, further theoretical
work \cite{EKN} - \cite{Langacker},
has shown that, once the condition of equality of all
supersymmetric particle masses is relaxed, the effective
supersymmetric threshold scale $T_{SUSY}$
can be very far from
any of the sparticle mass scales.
Thus, the meaning of $T_{SUSY}$ as the parameter which represents the
characteristic sparticle mass scale is no longer valid.
Nevertheless, as we shall explicitly show below,
the consequences of an arbitrary sparticle spectrum
on the predictions for the measurable low energy gauge couplings  can
always be  parametrize in terms of a single threshold scale
$T_{SUSY}$ \cite{Langacker}.

Due to the significance of the effective scale $T_{SUSY}$ for
our study, it is relevant to analyze its behavior
in detail.
Langacker and Polonsky
\cite{Langacker}  have derived a formula which determines $T_{SUSY}$
in terms of three low energy effective threshold scales $T_i$
associated with each gauge coupling $\alpha_i$ \footnote{ Observe that
there is a slight difference between our notation and that one used by
the authors of
Ref. \cite{Langacker}. The $M_i$ scales there are our $T_i$ and
they have used the parameter $M_{SUSY}$ to denote both the meaning of
our parameters $M_{SUSY}$ and $T_{SUSY}$.}.
However, in order to obtain more information from the dependence
of our predictions on the effective supersymmetric threshold scale,
 we find it most useful to present $T_{SUSY}$
in terms of the physical masses of the supersymmetric
particles.

\subsection{Threshold Correction Effects}

We assume
the unification of the three gauge couplings at $M_{GUT}$ and
consider the beta functions for their running up to two loops.
Thus,  after the inclusion of the one loop
threshold corrections coming from the supersymmetric particles,
for each gauge coupling $\alpha_i$, we obtain
the  relation
\begin{equation}
\frac{1}{\alpha_G} = \frac{1}{\alpha_i(M_Z)} -
\frac{b_i^{MSSM}}{2 \pi} \ln \left( \frac{M_{GUT}}{M_Z}
\right) + \gamma_i +
\frac{1}{\alpha_i^{Sthr}}  + \Delta_i,
\label{eq:alphag}
\end{equation}
 where $\gamma_i$ contains the two loop contributions to the beta
function, while
\begin{equation}
\frac{1}{\alpha_i^{Sthr}} =
\sum_{\eta, M_{\eta} > M_Z}
 \frac{b_i^{\eta}}{2\pi} \ln\left( \frac{M_{\eta}}
{M_Z} \right) ,
\label{eq:alit}
\end{equation}
are one loop
threshold corrections to $1/\alpha_i(M_Z)$.
In the above,
the summation is over  all
sparticle and heavy Higgs doublet states
with masses above $M_Z$,
$b_i^{\eta}$
is the contribution of each sparticle (or heavy Higgs)
to the one loop
beta function coefficient
of the gauge coupling $\alpha_i$ and $b_i^{MSSM}$
is  the one loop beta function coefficient
 of the gauge coupling
$\alpha_i$ in the MSSM.
Finally, the term $\Delta_i$ includes the  threshold
 corrections related to the top quark mass as well as a constant term
which depends only on the gauge group, $\Delta_1 = 0$,
$\Delta_2 = -1/6 \pi$,  $\Delta_3 = -1/4 \pi$,  and results from the
necessity of converting the gauge couplings, which are evaluated
in the $\bar{MS}$ scheme, to their
expressions in the Dimensional Reduction ($\bar{DR}$)
scheme, in order to work consistently in the MSSM
\cite{SCJN}.
It is important to remark that,
when solving the renormalization group equations at
the two loop  level, it is in general
sufficient to consider threshold
corrections at the one loop level. In the above, we have not
included threshold corrections arising at the grand unification
scale. The quantitative study of these corrections requires the
knowledge of the model describing the physical interactions at
the high energy scale, something that is beyond the scope of
the present analysis.
We shall elaborate on this issue in a future
work \cite{Prep}.

The unification condition allows to determine one coupling,
for example $\alpha_3(M_Z)$, as a function of the electroweak
gauge couplings  at $M_Z$
 and the one loop supersymmetric
 threshold corrections.
 Following first the parametrization of
Langacker and Polonsky \cite{Langacker},
let us define three low energy
effective threshold
scales
$T_i$, associated with the one loop
supersymmetric threshold corrections
to the gauge couplings $\alpha_i$ by
\begin{equation}
\sum_{\eta, M_{\eta}>M_Z}
\frac{b_i^{\eta}}{2\pi} \ln\left( \frac{M_{\eta}}
{M_Z} \right) \equiv \frac{b_i^{MSSM} - b_i^{SM}}{2 \pi}
\ln \left( \frac{ T_i}{M_Z} \right),
\label{eq:thi}
\end{equation}
where $b_i^{SM}$ is the one loop beta function coefficient
 of the gauge coupling
$\alpha_i$ in the  SM.
 Observe that,
with the above definition, Eq.(\ref{eq:thi}), if all sparticles
acquire masses above the scale $M_Z$ the effective scale
$T_i$ will be larger than $M_Z$ and of the order of the
characteristic mass of the sparticles
contributing to $b_i^{MSSM}$.
{}From Eqs.
(\ref{eq:alphag}) -
(\ref{eq:thi}),  it follows that,
\begin{eqnarray}
\frac{1}{\alpha_3(M_Z)}  &=&
\frac{ \left(
b_1 - b_3 \right)}
{ \left(
b_1 - b_2 \right)} \left[
\frac{1}{\alpha_2(M_Z)}
 + \gamma_2  + \Delta_2 \right]
-
\frac{ \left(
b_2 - b_3 \right)}
{ \left(
b_1 - b_2 \right)}  \left[
\frac{1}{\alpha_1(M_Z)}
 + \gamma_1  + \Delta_1 \right]
\nonumber\\
& - & \gamma_3  - \Delta_3
+ \Delta^{Sthr}\left(\frac{1}{\alpha_3(M_Z)}\right)
\label{eq:pred3}
\end{eqnarray}
where
\begin{equation}
\Delta^{Sthr} \left( \frac{1}{\alpha_3(M_Z)} \right) =
\frac{ \left(
b_1 - b_3 \right)}
{ \left(
b_1 - b_2 \right)}
\frac{1}{\alpha_2^{Sthr}}
-
\frac{ \left(
b_2 - b_3 \right)}
{ \left(
b_1 - b_2 \right)}
\frac{1}{\alpha_1^{Sthr}}
- \frac{1}{\alpha_3^{Sthr}},
\label{eq:alpha3t}
\end{equation}
is the contribution
to $1/\alpha_3(M_Z)$ due to the inclusion of the supersymmetric
threshold corrections at the one loop level.
In the above,
for simplicity of notation, we have dropped the superscription
from the supersymmetric beta functions, naming
$b_i^{MSSM} \equiv b_i$.
Replacing the values of  $b_i$ and $b_i^{SM}$
in Eq.(\ref{eq:alpha3t}), while using Eqs.(\ref{eq:alit})
and
(\ref{eq:thi}),
we obtain
\begin{equation}
\Delta^{Sthr}\left(\frac{1}{\alpha_3(M_Z)}\right) =
\frac{1}{28 \pi} \left[
100 \ln \left(\frac{T_2}{M_Z}\right)
-25 \ln \left(\frac{T_1}{M_Z}\right)
-56 \ln \left(\frac{T_3}{M_Z}\right)
\right].
\label{eq:talpha3}
\end{equation}
Thus, the supersymmetric threshold corrections,
Eq.(\ref{eq:alpha3t}) are characterized by the mass scales
$T_i$.

\subsection{The Effective SUSY Threshold Scale}

The effective threshold scale
 $T_{SUSY}$ is defined as that one which
would produce the same threshold correction given
in Eq.(\ref{eq:talpha3}), in the case in which
all the supersymmetric particles were degenerate in mass.
Hence, as a function of the mass scales $T_i$, it reads,
\begin{equation}
100 \ln\left( \frac{T_2}{M_Z} \right)
-25 \ln\left( \frac{T_1}{M_Z} \right)
-56 \ln\left( \frac{T_3}{M_Z} \right) =
19 \ln\left( \frac{T_{SUSY}}{M_Z} \right) ,
\label{eq:msusy}
\end{equation}
in agreement
 with the expression given by Langacker and
Polonsky \cite{Langacker}.
{}From equation (\ref{eq:msusy}) we can easily see that,
unless $T_1 \simeq T_2 \simeq T_3$, the effective  supersymmetric
threshold  scale $T_{SUSY}$ may
significantly
differ from the actual value of the masses $T_i$.
In particular, it may
be lower than $M_Z$.

The low energy input parameters for the RG equations relevant to
achieve gauge coupling unification are given at the scale $M_Z$. Thus,
it is necessary to understand, which is the role of $T_{SUSY}$ in the
case in which its value occurs to be below $M_Z$.
In fact, if,
 for a fixed set of values of the low energy coupling constants
$\alpha_i(M_Z)$, the unification of couplings takes place for a given
splitting of the
sparticle spectrum equivalent to consider
  a value of
 $T_{SUSY} > M_Z$, then,
 for the same values of  $\alpha_i(M_Z)$,
 other different splittings of the
   sparticle spectrum will still be compatible with
   gauge coupling unification,
as far as the value of the
effective threshold scale, as defined in
Eq.(\ref{eq:msusy}), is preserved.  In particular,
the case in which all sparticles acquire
a mass $M_{SUSY} = T_{SUSY}$
will be the possible
alternative for vanishing splitting  of the supersymmetric
 spectrum.
If $T_{SUSY} < M_Z$,
instead, for a given set of low energy parameters
the unification of the  gauge couplings at a scale $M_{GUT}$ will only
be achieved  in the case of
a sufficiently large sparticle mass splitting, which, through
Eq.(\ref{eq:msusy}), gives the corresponding low value for $T_{SUSY}$.
Moreover,
$T_{SUSY}$ conserves a proper mathematical meaning: Unification
of couplings will  be achieved if we run down
the standard model RG equations from $M_Z$ to $T_{SUSY}$ and
then we run up the supersymmetric RG equations from
$T_{SUSY}$ to the unification scale.
On the contrary,
if all sparticles had  a  characteristic mass $M_{SUSY}$
 lower than $M_Z$, they would
produce no threshold corrections to the values of $\alpha_i(M_Z)$.
Thus, in such case $T_{SUSY}$ can   not be but $M_Z$, and, therefore,
it is no longer possible to identify it
 with $M_{SUSY}$ as before.
 This means that,  for values
of the low energy couplings, for which we
obtain unification for a splitted supersymmetric spectrum which yields
a value of
$T_{SUSY} < M_Z$, we
would never obtain
unification in the case of a degenerate
supersymmetric spectrum.

A remark is in order. The scale $T_{SUSY}$ is only
related to the condition of unification and not to the
specific value of $M_{GUT}$ and $\alpha_G(M_{GUT})$.
In fact, for the same value of $T_{SUSY}$,  different
sparticle spectra  will produce the
unification of the gauge couplings at  different
grand unification scales and with  different values of the
coupling constant $\alpha_G(M_{GUT})$. This  has
implications for other constraints, coming, for
example, from the nonobservation of proton decay \cite{AN} -
\cite{LNPZ}.

\subsection{$T_{SUSY}$ and the Physical Sparticle Masses}

In order to study the dependence of $T_{SUSY}$ on the
different sparticle mass scales of the theory,
we define
$m_{\tilde{q}}$,
$m_{\tilde{g}}$,
$m_{\tilde{l}}$,
$m_{\tilde{W}}$,
$m_{\tilde{H}}$ and $m_H$
as the characteristic masses of the
squarks, gluinos, sleptons, electroweak gauginos,
Higgsinos and  the heavy Higgs doublet,
respectively. Assuming different values for all these
mass scales, we derive an expression for
the effective supersymmetric threshold $T_{SUSY}$ which is given
by
\begin{equation}
T_{SUSY} =
m_{\tilde{H}}
\left( \frac{m_{\tilde{W}}
}{m_{\tilde{g}}}
\right)^{28/19}
\left[
\left( \frac{m_{\tilde{l}}}{m_{\tilde{q}}}
\right)^{3/19}
\left( \frac{m_H}{m_{\tilde{H}}}
\right)^{3/19}
\left( \frac{m_{\tilde{W}}}{m_{\tilde{H}}}
\right)^{4/19} \right] .
\label{eq:susym}
\end{equation}
The above relation holds  whenever
 all particles considered above have a mass
 $m_{\eta} > M_Z$. If, instead,
 any of the sparticles or the heavy Higgs boson
has a mass $m_{\eta} < M_Z$,
it  should be replaced
by $M_Z$ for the purpose of computing the
supersymmetric threshold
corrections to $1/ \alpha_3(M_Z)$.
In the following,  unless otherwise
specified, we shall assume that
all sparticles
and the heavy Higgs doublet acquire masses above $M_Z$.
{}From Eq.(\ref{eq:susym}), it follows that,
for fixed mass values of the uncolored sparticles,
that is sleptons, Higgsinos
and the weak gauginos, together with the heavy Higgs doublet,
the value of $T_{SUSY}$ decreases for larger mass
values of the colored
sparticles -
 squarks and
gluinos. Moreover,
$T_{SUSY}$ depends
strongly on the first two factors in Eq.(\ref{eq:susym}),
while it
is only slightly
dependent on the expression
 inside the squared brackets. This is most surprising, since it
implies that $T_{SUSY}$ has only a slight dependence on the
squark, slepton and heavy Higgs
masses and a very strong dependence on
the overall Higgsino mass, as well as on the ratio of
masses of the gauginos associated with the
electroweak and strong interactions. The mild dependence
of the supersymmetric threshold corrections on the squark and
slepton mass scales is in agreement with a similar observation
made in the context of the  minimal supersymmetric SU(5) model
in Ref.\cite{EKN}.

In the above, we have implicitly identified the sparticle
mass eigenstates with $SU(3)_c\times SU(2)_L\times U(1)_Y$
eigenstates. Since the most significant dependence of
$T_{SUSY}$ on the supersymmetric particle masses comes from
the gaugino and Higgsino sectors,
we expect no significant dependence of $T_{SUSY}$
on the squark and slepton mixing. The
neutralino and Higgsino mixing may,
instead,
have quantitative effects, which have not been included
above. We shall deepen into this point elsewhere \cite{Prep}.

In some models, in which
the source of supersymmetry breaking is given by a
common gaugino mass, $M_{1/2}$, at high
energy scales, it follows that,
within a good approximation,
all colored sparticles
have a common mass $m_{col}  = {\cal{O}}(m_{\tilde{g}})$
while all uncolored
sparticles  acquire a common mass $m_{wk} = {\cal{O}}(M_{1/2})$.
The mass of
the Higgsinos is determined by the supersymmetric mass
parameter $\mu$ appearing in the superpotential and we are
implicitly assuming that $|\mu| = {\cal{O}}(M_{1/2})$.
 Thus, in the  simplified case
 in which only two different supersymmetric mass
scales are presumed,
the effective  supersymmetric threshold
scale is given as a function of
both mass parameters $m_{col} $ and
$m_{wk}$ as follows,
\begin{equation}
T_{SUSY} = m_{wk} \left( \frac{m_{wk}}
{m_{col}} \right)^{31/19}.
\label{eq:tsusy}
\end{equation}
As we remarked above for the most general case, if any
one of these two masses is below $M_Z$, it
should be identified with $M_Z$ while applying Eq.
(\ref{eq:tsusy}).
Assuming $m_{col} > m_{wk}$, $T_{SUSY}$ is
predicted to be smaller than $m_{wk}$. Moreover, since
$T_{SUSY}$ decreases for larger values of $m_{col}$ we have
that for a fixed value of $m_{wk}$, the larger the splitting
of masses the smaller becomes $ T_{SUSY}$.
For example, if all the colored sparticles are assumed
to have masses of the order of
 $m_{col} \simeq 1$ TeV, while the uncolored ones
have masses of the order of
$m_{wk} = 200$ GeV, the resulting effective
supersymmetric threshold scale is as low as
$T_{SUSY} \simeq 15$ GeV.
Of course, the fact of considering all colored
sparticles to be degenerate in mass
and the same for all the uncolored ones
is  an approximation, which becomes natural
only in some particular supersymmetry breaking schemes.
However,
due to the slight dependence of $T_{SUSY}$
on the squark, slepton and heavy Higgs masses, the  effective
supersymmetric threshold scale given in
Eq.(\ref{eq:susym}) is
approximately described  by
\begin{equation}
T_{SUSY} \simeq m_{\tilde{H}} \left( \frac{m_{\tilde{W}}}{
m_{\tilde{g}}} \right)^{3/2}.
\label{eq:appsusy}
\end{equation}
Therefore, the approximation
made above, Eq. (\ref{eq:tsusy}),
turns out to be a rather good one if one identifies
$m_{col}$ with the gluino mass, while
$m_{wk}$  is a geometrical average of the
overall value
$m_{\tilde{H}}$ of the Higgsinos and that one of the
electroweak
gauginos.
 This simplification yields a
clearer computational framework for all quantities
 with a dominant dependence on $T_{SUSY}$.

Depending on the SUSY breaking scheme, it may occur that
there is no mass splitting between squarks and sleptons.
If, for example, the dominant source of soft supersymmetry
breakdown is a common
mass $m_0^2 \gg M_Z^2$ for the spin-0 sparticles,
while the common gaugino mass $M_{1/2}$ is of the order of $M_Z$,
then
squarks, sleptons and  the heavy Higgs particles
will  become heavy and almost degenerate
in mass, while gauginos will be relatively light \cite{KLNQ}.
However, in such framework as well as in the case in which
squarks and sleptons are far from being degenerate in mass,
the behavior of the supersymmetric threshold scale
as a function of the sparticle
masses implies that Eq.(\ref{eq:appsusy}) gives, to a good
approximation, the value of $T_{SUSY}$.

In  many grand unified scenarios,
the supersymmetric
spectrum is determined by the soft supersymmetry breaking
parameters
arising at $M_{GUT}$, that is
a common scalar mass $m_0$, a common gaugino mass $M_{1/2}$,
and the scalar trilinear couplings
 $A_0$ and $B_0$ involving
 contributions proportional to the
trilinear and bilinear terms in the superpotential, as well as
the supersymmetric mass parameter $\mu_0$, appearing in the
superpotential. Moreover the assumption of
 small gaugino - Higgsino mixing implies that
the overall
Higgsino mass is approximately given by the absolute value of
the renormalized
supersymmetric mass $\mu$, while, the assumption of an
overall common gaugino mass at $M_{GUT}$ implies that
the ratio of the gluino to the wino mass is approximately
given by the ratio of the strong to the weak gauge coupling.
Therefore, if all sparticle masses are   above $M_Z$,
we obtain that the approximate expression
of $T_{SUSY}$, Eq.(\ref{eq:appsusy})  simplifies to
\begin{equation}
T_{SUSY} \simeq
 m_{\tilde{H}} \left( \frac{\alpha_2(M_Z)}{\alpha_3(M_Z)}
\right)^{3/2}.
\label{eq:apptsu}
\end{equation}
Thus, the effective supersymmetric threshold
scale is proportional to the absolute value of the
supersymmetric mass $\mu$
($T_{SUSY} \simeq |\mu|/7$), and it is by no means a measure of
the scale of  soft supersymmetry breaking. Taking
values for $|\mu| < 1$ TeV, we obtain that the effective
threshold scale is of the order of, or smaller than $M_Z$.
Observe that, since within this framework the threshold
corrections to $\alpha_3(M_Z)$ depend on $\alpha_3(M_Z)$
itself, the value of the strong gauge coupling
must be found by solving
Eq.(\ref{eq:pred3}) in a self consistent way.
Within the minimal SU(5) supersymmetric
model, this was performed in Ref.\cite{EKN}.

Once more, if
any of the  sparticles acquires a mass which is small
in comparison with $M_Z$, this mass must be
replaced by $M_Z$ when applying Eq.(\ref{eq:appsusy}).
For example, if the supersymmetry breaking parameter
$M_{1/2}$ were  very small in comparison with $M_Z$,
all gaugino masses would be smaller than
or of the order of
$M_Z$. In this case, the effective supersymmetric threshold
scale
would be approximately given by the Higgsino mass. This
situation is, however, not favored by experiments,
apart from the possible existence of a
light gluino window \cite{Glu}. Another
possibility is that the winos acquire masses
below $M_Z$, while the gluinos are heavier than the $Z^0$
vector
neutral boson.  In such case, one should replace the wino
mass by $M_Z$ directly in Eq.(\ref{eq:appsusy}).

\subsection{Numerical Computations}

  Once the low energy values of the fine structure constant,
$\alpha(M_Z)$, and $\sin^2 \theta_W(M_Z)$ are fixed, the
condition of gauge coupling unification allows to
determine the value of $\alpha_3(M_Z)$, up to threshold
corrections. For the aim of performing the numerical
computations, rather than including the supersymmetric threshold
corrections through the procedure  explained above,
we shall approach the issue in
the following equivalent way:
At energies above the heaviest sparticle mass,
we  use the complete two loop renormalization group
equations of the MSSM, including the effect of the $SU(3)_c
\times SU(2)_L \times U(1)_Y$  gauge couplings and the third
generation of quark  and lepton  Yukawa couplings. Below the scale
of energy given by each sparticle mass, we account for the
threshold corrections by modifying the one loop renormalization
group equations
while keeping the two loop contributions as in the complete
MSSM (this is a correct procedure, up to higher order corrections).
We shall assume, as in Eq.(\ref{eq:tsusy}), that all colored
sparticles acquire a common mass $m_{col}$, while the uncolored
ones, together with the heavy Higgs doublet acquire a mass
$m_{wk}$.
Thus, in practice, we consider a two
step decoupling for the supersymmetric particles and the heaviest
Higgs doublet.
 At energy scales $ m_{wk} \leq \mu
\leq m_{col}$, the model has the same particle content
as in the standard model with two Higgs doublets and with the
addition of all the uncolored sparticles.
At energies
below $m_{wk}$, we recover the standard model particle
spectrum.
Thus, we account for
the uncolored sparticle  and heavy Higgs doublet
threshold corrections while decoupling them
at  $m_{wk}$
 and   using
the full two loop renormalization group equations of the
SM for the energy range
$ M_Z \leq \mu
\leq m_{wk}$.
 Furthermore,
since the top quark mass fulfills the relation $M_t \geq M_Z$,
the top quark threshold corrections must be also taken
into account. We include
the top quark threshold corrections to the
gauge couplings analytically,
by the procedure explained in Ref.\cite{Langacker}.

The inclusion  of the Yukawa sector in our study requires extra boundary
conditions for our numerical analysis, besides the low energy
 electroweak parameters and the condition of gauge coupling unification
already implemented  in the gauge sector. Indeed,
 the values of the bottom quark and tau masses
 are necessary input
data for the low energy boundary conditions while the unification
of their Yukawa couplings determines the boundary conditions at the
unification scale. Then, we obtain predictions for the top quark mass as
a function of $\tan \beta$.
 In the extreme case  in which we  require the
unification of the Yukawa couplings of the top and bottom quarks
and the tau lepton at the unification scale -
$h_t(M_{GUT}) = h_b(M_{GUT}) = h_{\tau}(M_{GUT}) $ -
a prediction for $\tan \beta$ is derived as well.
Obviously, while performing the RG evolution in the Yukawa  sector,
in order to be consistent within our framework, we have to
consider the same two step decoupling procedure at
$m_{col}$ and  $m_{wk}$ and to include the threshold
corrections associated with the decoupling of the top quark, too.
At energy scales
$\mu < M_Z$, we have an effective $SU(3)_c \times U(1)_{em}$
symmetric theory. Hence,
we consider
the running of the bottom quark and $\tau$ lepton  masses
within the SU(3)$_c$ $\times$ U(1)$_{em}$ gauge group.

\subsection{The Higgs Sector}

We also consider the Higgs sector of the theory \cite{OYY} -
\cite{HH2}.
 At
energy scales   above $m_{col}$,
the Higgs potential in the MSSM
with  general soft-SUSY breaking
terms is given by \cite{Dyn} - \cite{HH2},
\begin{eqnarray}
V_{\rm eff}&=m_1^2H_1^{\dagger}H_1+
m_2^2H_2^{\dagger}H_2
-m_3^2(H_1^Ti\tau_2H_2+h.c.)
 \nonumber\\
&+{1\over8}(g_1^2+g_2^2)(H_2^{\dagger}H_2-H_1^{\dagger}H_1)^2
+{1\over2}{g_2^2}\vert H_2^{\dagger}H_1\vert ^2,
\end{eqnarray}
where $g_2$ and $g_1$ are the gauge couplings
of $SU(2)_L$ and
$U(1)$, respectively. The mass parameters $m_{i}$, with $i=1,2,3$,
 as well as the gauge
couplings are running parameters depending
on the renormalization
scale.
  The parameter $M^2$, defined as $m_1^2+m_2^2=2M^2$, must be
of the order of
$m_{wk}$, to be compatible with the assumption that only one light
Higgs doublet appears in the low energy spectrum of the theory.
 At energy
scales $ \mu
\leq m_{col}$,
the potential
is given by the general expression
\begin{eqnarray}
V_{\rm eff}&=&m_1^2H_1^{\dagger}H_1+
m_2^2H_2^{\dagger}H_2
-m_3^2(H_1^Ti\tau_2H_2+h.c.)
\nonumber\\
&+&{\lambda_1\over2}(H_1^{\dagger}H_1)^2+{\lambda_2\over2}
(H_2^{\dagger}H_2)^2
+{\lambda_3}(H_1^{\dagger}H_1)(H_2^{\dagger}H_2)
+{\lambda_4}\vert H_2^{\dagger}i\tau_2H_1^{\ast}\vert ^2.
\label{eq:Higgsp}
\end{eqnarray}
Therefore, the
 running quartic couplings $\lambda_{j}$, with $j=1-4$,
must satisfy the following
boundary conditions at  $m_{col}$:
\begin{equation}
\lambda_1 = \lambda_2 ={{g_1^2+g_2^2}\over4}, \qquad
\lambda_3
={{g_2^2-g_1^2} \over 4},\qquad \lambda_4 = -{{g_2^2}\over2},
\end{equation}
where we have used the relation
\begin{equation}
(H_1^{\dagger}H_1)(H_2^{\dagger}H_2)=
\vert H_2^{\dagger}H_1\vert ^2
+\vert H_2^{\dagger}i\tau_2H_1^{\ast}\vert ^2.
\end{equation}
At  energy scales $\mu$ below $m_{col}$,
 the values
of the quartic couplings may be obtained
by solving the corresponding renormalization group
equations \cite{HH2},\cite{CH},
considering the step decoupling  described above.
As we explained in Ref.\cite{Dyn},
the appearence of other quartic
couplings than those ones given in
Eq.(\ref{eq:Higgsp}) is protected by either
discrete symmetries ($H_2 \rightarrow -H_2$; $H_1 \rightarrow -H_1$)
or a global PQ symmetry. In particular, a quartic term
$\lambda_5 [(H_2^T i \tau_2 H_1)^2 + h.c.] $
breaks the PQ symmetry.
The PQ symmetry is only broken by mass or scalar trilinear
terms, for example the term $m_3^2$ appearing in the
scalar potential, Eq. (\ref{eq:Higgsp}).
Hence, $\lambda_5$ will not receive any leading
logarithmic contribution and will not appear in the renormalization
group analysis.

Quite generally, after the electroweak symmetry breaking,
three neutral and two charged scalar
states appear in the spectrum. As we said above, in our
analysis we are considering the case in which only
one light Higgs doublet $\Phi$, which is a combination
of the original Higgs doublets $H_1$ and $H_2$, remains
below $m_{wk}$. It is straightforward to prove
that the mass matrices of the charged and CP-odd states are
always diagonalized with an $H_1-H_2$ mixing angle which
is given by $-\beta$. In addition, whenever the mass
parameter $M \gg M_Z$, the mixing angle for the neutral CP-even
states is approximately given by $\beta$ and, hence, the
lightest CP-even state, together with the Goldstone modes
form a Higgs doublet $\Phi$, whose expression is given by
\begin{equation}
\Phi = H_1 \cos\beta + i \tau_2 H_2^{*} \sin\beta.
\end{equation}
The mass of the lightest CP even Higgs particle is given
by
\begin{equation}
m_h^2 = 2 \lambda(m_t) v^2
\label{eq:mhiggs}
\end{equation}
where $v^2 = v_1^2 +v_2^2$, with $v_{1,2}$ the vacuum
expectation values of the Higgs fields $H_{1,2}$, respectively,
and $\lambda(\mu)$ is the effective quartic coupling of the
light Higgs doublet.
The low energy value of the quartic coupling
$\lambda$ may be obtained by using the RG equations with
the boundary condition
\begin{equation}
\lambda = \lambda_1 \cos^4\beta + \lambda_2 \sin^4 \beta
+ 2 (\lambda_3 + \lambda_4) \cos^2\beta \sin^2\beta
\end{equation}
at the scale $\mu = m_{wk}$.
In Eq.(\ref{eq:mhiggs}) we have defined
the light Higgs running mass at $m_t$, which, in a very good
approximation coincides with the on shell mass $m_h(m_h)$.
This is due to the fact that the only relevant
radiative corrections to the Higgs particle
are induced by the top quark, and in the MSSM,
for values of the top quark mass compatible with experimental
bounds,
$m_t > M_Z$, the relation    $m_t \geq m_h$ is fulfilled.
Therefore, the evaluation of the light Higgs
running mass at $m_t$ allows a correct
evaluation of the top quark threshold corrections, while,
below $m_t$ the Higgs mass running is negligible.

The full two loop RG equations we
use in the MSSM and the
SM regimes, are given in Ref.\cite{RG} and Refs.\cite{RSM},
\cite{RGSM}, respectively.
In addition,
the RG equations for the bottom quark and tau lepton considered at
scales $\mu \leq M_Z$, are given, for example, in Ref.\cite{RSM}.
The proper one loop
renormalization group equations, describing
the evolution of the gauge, Yukawa and
quartic  couplings, in the case in which
the uncolored sparticles  are the only supersymmetric
contributions to
the particle  spectrum, as it occurs at the
intermediate scales $m_{col} \leq \mu \leq m_{wk}$, are
given in Refs.\cite{HH2} , \cite{CH}.
We solve the RG equations numerically using the above described
multiple step decoupling procedure.

\section{Unification of Couplings}

As we mentioned in the introduction,
there have been several recent studies dealing with the
problem of gauge and Yukawa coupling unification within
the MSSM. One of the
most interesting results is that the values of the gauge
couplings obtained from the most recent experimental
data is consistent with the unification of gauge couplings
within the MSSM, if the supersymmetric threshold scale is
of the order of the weak scale. Although, quite generally,
this statement is correct,
there are several important points which should be
taken into account. In what follows,
we shall  assume that the unification of
the weak and strong
gauge couplings takes
place, and we shall leave the inclusion of
threshold corrections arising at the GUT
scale for a future analysis \cite{Prep}. In such context,
the following properties are fulfilled:
{}~\\
I) For a given value of $T_{SUSY}$, the predicted value of
$\alpha_3(M_Z)$ depends strongly on the exact value of
$\sin^2 \theta_W(M_Z)$, and increases (decreases)
when $\sin^2
\theta_W(M_Z)$ decreases (increases). \\
II) As can be seen from Eqs.(\ref{eq:pred3})-(\ref{eq:msusy}),
for a given value of $\sin^2 \theta_W(M_Z)$, the
predicted value of $\alpha_3(M_Z)$ is only logarithmically
dependent on $T_{SUSY}$ and increases (decreases)
when $T_{SUSY}$ decreases (increases).  \\
III) At the two loop level, the contributions of the top
quark Yukawa coupling to  $\alpha_3(M_Z)$  become relevant.
For example, for the central values
 $\sin^2 \theta_W(M_Z) = 0.2324$, $M_t = 138$ GeV,
$T_{SUSY} \simeq M_Z$ and $\tan\beta$ such that the
top quark Yukawa coupling $h_t(m_t) \approx h_t(M_{GUT})
\approx 1$,
the predicted value of the strong gauge coupling reads
$\alpha_3(M_Z) \simeq 0.124$ \cite{Langacker}. However,
since at the
two loop level the strong gauge coupling receives negative
contributions
from the top quark Yukawa coupling, a larger value of the
top quark Yukawa coupling at $M_{GUT}$ will imply a lower
value for $\alpha_3(M_Z)$. In fact, $\alpha_3(M_Z)$
may differ in up to a
2$\%$ while varying the top quark Yukawa coupling from
$h_t(M_{GUT}) \simeq 1$ to
very large values, $h_t^2(M_{GUT})/4\pi \approx 1$.

To analyze the consequences of properties I)-III), one must
remember that, as we discussed in  section 3,
the effective scale
$T_{SUSY}$ is directly related to the overall
scale of sparticles masses only in the unlikely case of
an absolute
degenerate supersymmetric spectrum. If a significant mass
splitting  is allowed, $T_{SUSY}$ can
significantly differ from the characteristic sparticle
mass scale. In particular, even when all  sparticle masses are
larger than $M_Z$, the effective $T_{SUSY}$ may be well
below the $Z^0$ mass scale. Indeed, for a given value
of $\sin^2 \theta_W (M_Z)$, property II) implies that
 it is possible to
 obtain a lower bound
on $T_{SUSY}$ by requiring $\alpha_3(M_Z)$ to be below
its present experimental upper bound
\begin{equation}
\alpha_3(M_Z) \leq 0.13 .
\label{eq:alpha3c}
\end{equation}
In principle, one could obtain an upper bound
on $T_{SUSY}$ from the experimental lower bound on the
strong gauge coupling ($\alpha_3(M_Z) \geq 0.11$), as well.
However, in the present analysis, in which no threshold
corrections at the unification scale
are included, the value of $T_{SUSY}$ needed to obtain
values of $\alpha_3(M_Z)$ below its experimental value
is, in general, of the order of or greater than 1 TeV
and, hence, it does not yield significant constraints on the
sparticle spectrum, once the absence of fine tunning is
required. For  example, for $\sin^2 \theta_W(M_Z)
= 0.2335$ and $T_{SUSY} = 1$ TeV, we obtain
$\alpha_3(M_Z) \simeq 0.111$. Thus, the constraint
$T_{SUSY} \leq
{\cal{O}}$(1 TeV) follows in this case. For smaller
values of $\sin^2 \theta_W(M_Z)$ the constraints
on $T_{SUSY}$ are even weaker.

For  given low energy
values of the gauge couplings, the requirement
of bottom and tau Yukawa coupling unification determines
the values of the top quark mass
as a function of $\tan \beta$, depending on the input value
of the bottom quark mass \cite{Ramond}-\cite{DHR}.
It has been observed \cite{Ramond},\cite{Dyn}
that, for running bottom quark masses $m_b(M_b)$ lower or
equal to 4.6 GeV, the top quark mass obtained in these
analyses is remarkably close to its infrared quasi
fixed point value \cite{IR}
associated with the \lq\lq triviality" bound
on the top quark Yukawa coupling.  As a matter of fact,
once the QCD
corrections are taken into account, from Eq.(\ref{eq:botmass}),
it is clear that for
values of the bottom quark mass consistent with the
experimental observations, $M_b \leq 5.2$ GeV, the
running bottom mass is bounded to be $m_b(M_b)
 \leq 4.6$ GeV. Therefore,
the top quark
Yukawa coupling obtained in the context of bottom and  tau Yukawa
couplings unification is, indeed, remarkably
 close to its infrared quasi fixed
point value. Moreover,
a value of the top quark Yukawa coupling close to its
infrared quasi fixed point necessarily implies that it
is getting large at scales of the
order of the grand unification scale \cite{Dyn}.

\subsection{Infrared Quasi Fixed Point Predictions}

The results associated with the infrared (IR)
 quasi fixed point can be readily
obtained analytically. As a matter of fact, recalling the
expressions for the beta  functions of the top and bottom
quark Yukawa couplings,
and
neglecting the small contributions of the weak couplings
and the tau lepton in this approximation,
we observe that
 the existence
 of the IR quasi fixed point \cite{Dyn},\cite{IR}
 implies
straightforward relations among $\alpha_3 $ and
$Y_t=h_t^2/(4\pi)$ and $Y_b=h_b^2/(4\pi)$,
at an energy scale of order $m_t$.
Namely, from the one loop beta function of
the top quark Yukawa coupling we obtain,
\begin{equation}
(16/3) \alpha_3(m_t) \simeq 6 Y_t(m_t) + Y_b(m_t).
\label{eq:IR}
\end{equation}
Moreover, the bottom quark
Yukawa coupling at the top quark
mass scale is approximately given by
\begin{equation}
h_b(m_t) \simeq \sqrt{1 +
{\tan \beta}^2} \;m_b(m_t)/v.
\end{equation}
Since $m_b(m_t) \simeq 3$ GeV,
for moderate values of $\tan \beta$, it follows that
 $Y_b(m_t) \ll
\alpha_3(m_t)$
and, hence, the bottom quark Yukawa coupling
contribution to Eq.(\ref{eq:IR}) can be safely neglected.
This implies that, for small
and moderate values of $\tan\beta \leq 30$,
using Eq.(\ref{eq:IR}) we can determine the
value of the top quark
Yukawa coupling as a function of the strong
gauge coupling constant. In addition, using the relation
\begin{equation}
m_t(m_t) \simeq h_t(m_t) \frac{v \tan\beta}{\sqrt{1 +
\tan^2\beta}},
\end{equation}
for a given value of $\alpha_3(m_t)$
we obtain the value of the top quark mass only as a function of
$\tan \beta $. For  larger values of $\tan \beta
\geq 30$, instead, the bottom
quark Yukawa coupling contribution to Eq.(\ref{eq:IR}) can no
longer be neglected. In fact, for a fixed value of $\alpha_3(M_Z)$
the value of the top quark Yukawa coupling decreases
with increasing $\tan \beta$  till, for a sufficiently large
value of $\tan\beta$, it becomes
equal to the bottom quark one. Then, the IR quasi fixed point relation
reads,
\begin{equation}
(16/3) \alpha_3(m_t) \simeq 7 Y_t(m_t) = 7 Y_b(m_t) .
\end{equation}
 For a given value of the strong gauge coupling and the
 bottom quark mass, we can determine the value
of $\tan\beta$ and hence of the top quark mass, for which the
above relation is fulfilled.
 Finally, for even larger values of $\tan\beta$,
we are in the regime in which $Y_b \gg Y_t$, and
the IR quasi fixed point relation
\begin{equation}
(16/3) \alpha_3(M_Z) \simeq 6 Y_b(M_Z),
\label{eq:maxb}
\end{equation}
holds. Since, for a
cutoff scale $\Lambda \simeq {\cal{O}}(10^{16})$ GeV,
the quasi infrared fixed point is associated with the
renormalization group trayectories corresponding to the
triviality bounds on the Yukawa couplings,
from Eq.(\ref{eq:maxb})
we obtain
a maximun allowed value for $\tan \beta$ , above which, a perturbative
treatment of the bottom quark Yukawa sector becomes inconsistent.

In Tables 1.a - 1.c,
we present the analytic IR quasi fixed point predictions
together with those obtained
by requiring the unification of the bottom and tau Yukawa couplings
within our two loop RG analysis, considering
$\sin^2\theta_W(M_Z) = 0.2324$,
$\alpha_3(M_Z) \simeq 0.122$ and  three different values of the
running bottom quark mass :
$m_b(M_b) \simeq 4.6, 4.3$ and 4.1 GeV.
For these values of the input parameters we
compare the results of both methods for the
values of the top quark mass
for three different values of $\tan \beta$
as well as for the value of  $\tan \beta$ and the top quark mass at
which the three Yukawa couplings of the third generation unify.
{}From the above comparison it follows that,
for moderate values of $\tan\beta$,
the top quark mass evaluated through the numerical RG approach
is only  two to four percent lower than its value
derived from the analytic approximation performed in the
context of the IR quasi fixed point.
Indeed, the values of $m_t$
get closer to  the  infrared quasi fixed point predictions (IRP)
for  lower  values of the running bottom quark mass.
As we said  before, the closeness of the
top quark mass predictions is related to the
relatively large values
of the top quark Yukawa coupling
$h_t$ at the grand unification scale. For very large values
of $\tan\beta > 40$ and larger values of $m_b(M_b)$,
the departure of the RG results
from the IRP is
larger, reflecting the lower values of the top quark Yukawa
coupling at the grand unification scale. For example, for the
three values of $m_b(M_b) = 4.6,4.3$ and 4.1 GeV, the
unification of the three Yukawa couplings is achieved for
$Y_t(M_{GUT}) \simeq 0.04, 0.13$ and 0.43 respectively.
Observe that, in the first case, the Yukawa couplings acquire
values which are of the order of the gauge coupling
value $\alpha_G(M_{GUT}) \simeq 0.04$.

\subsection{On the Perturbative Consistency of the Top Quark
Yukawa Sector}

For a consistent perturbative treatment of the theory, the top
quark Yukawa coupling must fulfill the condition $h_t^2(\mu)/
4 \pi \leq 1$ in the whole energy range in which the MSSM is considered
as a valid effective theory.
However, due to the existence of the IR quasi fixed point and the
behavior  of the top quark Yukawa coupling for intermediate
energy scales, the requirement of perturbative consistency is
naturally fulfilled  in the low to intermediate energy regime.
Thus, what has to be assured is the correct behavior of the top
quark Yukawa coupling in the high energy regime.
Since we are assuming that the
MSSM is valid up to scales of the order of $M_{GUT}$, an
upper bound, the so called triviality
bound on the above quantity
 may be defined as its  low energy
value  consistent with the renormalization
group trajectory for which it becomes strong at scales of
order $M_{GUT}$,
\begin{equation}
\frac{h_t^2 (M_{GUT})}{4 \pi} \leq 1.
\label{eq:Yukc}
\end{equation}
 This is equivalent to require that the
Landau singularity associated with this coupling occurs at
an energy scale $\mu_{sing} > M_{GUT}$.
As we shall analyze
in detail in section 6,
 the triviality bound
 on the top quark Yukawa coupling
 $h_t$ determines an upper bound
on $m_t$ as a function of  $\tan\beta$, which, for
values of $M_{GUT} \approx 10^{16}$ GeV, is approximately
given by the  infrared quasi fixed predictions for this quantity.

As a matter of fact, since
 the value of the top quark Yukawa coupling obtained within the
framework of bottom and tau Yukawa coupling  unification may
become quite large,  the requirement
of perturbative consistency of the  Yukawa sector of the
theory, Eq.(\ref{eq:Yukc}),
has important implications in defining bounds on the effective
supersymmetric threshold scale $T_{SUSY}$. To understand this,
we should emphasize  that the properties mentioned
at the beginning of  section 4, implying
that   the value of $\alpha_3(M_Z)$ increases (decreases) for
smaller (larger) values of $\sin^2 \theta_W$ as well as for
smaller (larger)
values of $T_{SUSY}$,  are independent of
the top quark
Yukawa sector and, hence, are also fulfilled  after
the inclusion
of the Yukawa coupling unification condition. In table 2,
we present a particular
set of predictions to support this general result.

A second point to be stressed is that  the
interrelation between the top quark Yukawa coupling and $\alpha_3$
is not only a property of the infrared, as shown in section 4.1,
 but,
 there is a strong
correlation between $Y_t(M_{GUT})$ and $\alpha_3(M_Z)$ as well.
Our analysis shows that, for values of the squark masses, which
do not involve  a large fine tuning,
$m_{\tilde{q}} \leq $ 3 TeV,
the actual value
of $Y_t(M_{GUT})$ strongly depends  on
the value of $\alpha_3(M_Z)$.
Indeed, once the values of the running bottom quark mass and
$\sin^2 \theta_W$
are fixed in each running, the value of the top quark
Yukawa coupling at $M_{GUT}$
 depends only on $T_{SUSY}$ and on
 $\alpha_3(M_Z)$. More specifically, $Y_t(M_{GUT})$
 increases (decreases) with
 $\alpha_3(M_Z)$, and, therefore,   it increases (decreases) for smaller
(larger) values of  $T_{SUSY}$ together with $\alpha_3(M_Z)$. We
illustrate this behavior in table 3. Moreover,
for a given $T_{SUSY}$,
the value of $Y_t(M_{GUT})$
is  almost independent of the spectific values of the
supersymmetric particle masses.
We illustrate this behavior  in Table 4,
where we fix the value of $T_{SUSY} = 1$ TeV and show the
slight variation of $Y_t(M_{GUT})$, $\alpha_3(M_Z)$ and
$m_{t}$ ($\tan\beta = 4$)
for $m_{col}$ varying from 1 to 10 TeV.
As readily seen from Table 4, the
radiative corrections to the Higgs particle mass, instead,
depend not only on the top quark Yukawa sector, and hence
on $T_{SUSY}$, but they have a logarithmic
dependence on the squark mass as well.
Most interesting is the fact that,
  due to the behavior of $\alpha_3(M_Z)$  and
$Y_t(M_{GUT})$ with
  the effective supersymmetric threshold scale, a lower bound on
it, $T^{min}_{SUSY}$, may be derived
 from the upper bound either  on
$\alpha_3(M_Z)$, Eq.(\ref{eq:alpha3c}), or on $Y_t(M_{GUT})$,
Eq.(\ref{eq:Yukc}).
A detailed analysis of these constraints is
presented in section 5.

\section{Constraints on the Splitting of the
Supersymmetric Particle Spectrum }

In this section we shall perform a quantitative analysis
of the constraints on the supersymmetric threshold scale,
which may be obtained from the experimental constraints on
 $\alpha_3(M_Z)$, Eq.(\ref{eq:alpha3c}), or from the
requirement of perturbative consistency of the top quark
Yukawa
sector of the theory, Eq.(\ref{eq:Yukc}).  These
constraints
have a strong dependence on the exact value of the low
energy parameters of the theory, in particular, of the bottom
quark mass
$M_b$ and $\sin^2 \theta_W(M_Z)$.
Concerning the bottom quark sector,
we study the cases $m_b(M_b) = 4.1, 4.3, 4.6$ GeV, which,
considering the proper two loop corrections
for the central value of $\alpha_3(M_Z) \simeq 0.12$,
approximately correspond to
a physical mass $M_b \simeq 4.7, 4.9, 5.2$ GeV, respectively.
The RG equations depend dominantly
on the running mass, having
a very
slight dependence on the physical mass  through the
point at which the running mass is evaluated. Hence,
for larger (smaller) values of $\alpha_3(M_Z)$, our results will
correspond to slightly larger (smaller) values of the physical
mass $M_b$, which are easily obtainable from
Eq.(\ref{eq:botmass}).

Although highly more accurate than the bottom quark
mass prediction,
the experimental prediction for  $\sin^2 \theta_W(M_Z)$,
Eq.(\ref{eq:sinmt}), depends
quadratically on the top quark mass. A way to deal
with this difficulty is to take the value of
$\sin^2 \theta_W(M_Z)$, which provides the best fit to the
experimental data with a free top quark mass value.
Consequently, we have considered that the weak mixing angle
takes values in the range
\begin{equation}
\sin^2 \theta_W(M_Z) = 0.2315 - 0.2335,
\end{equation}
which approximately corresponds to the range of values
obtained by such fit at the two sigma level\cite{Langacker}.
Even though  we shall not use the relation given
in Eq.(\ref{eq:sinmt}) to correlate the value of the top
quark mass and the weak mixing angle, we should mention
that the best experimental fit to the top quark mass
corresponding to the above values of the weak mixing angle
are  $M_t \simeq 170 - 100$ GeV, respectively.

To perform our analysis
we  proceed as follows: For fixed values of $\alpha(M_Z)$,
$\sin^2 \theta_W(M_Z)$, $m_b(M_b)$ and $m_{\tau}(M_{\tau})$
and a given effective supersymmetric threshold scale
$T_{SUSY}$, we  require the unification of the gauge couplings
 and the
bottom quark and tau Yukawa couplings to determine
$\alpha_3(M_Z)$ and the top quark Yukawa coupling $h_t(m_t)$,
respectively. We use the latter, $h_t(m_t)$, to predict the
on shell value of the top quark running mass $m_t(m_t)$
as a function of $\tan\beta$. It
is worth remarking  that, for moderate values of
$\tan \beta \leq 30$, the value of the top quark
Yukawa coupling obtained
from the unification condition has only a slight dependence
on this quantity and, hence, the variation of the top
quark mass with $\tan \beta$ arises almost exclusively from
the variation of the vacuum expectation value
of the scalar Higgs $H_2$, which couples to the
top quark. In this section, we shall implicitly assume
that the condition $\tan\beta < 30$ is fulfilled.

Using the renormalization group equations, we vary the
scale $T_{SUSY}$ until one of the two conditions, Eq.
(\ref{eq:alpha3c}) or Eq.(\ref{eq:Yukc}) is fulfilled.
This determines a lower bound on $T_{SUSY}$ below which,
gauge and Yukawa coupling unification lead to either
a loss of perturbative consistency of the top
quark Yukawa
sector, or to values of $\alpha_3(M_Z)$ which are
too large to be consistent with the
present experimental constraints on this quantity.
In Tables 5.a - 5.c we present the constraints on
the value of the effective supersymmetric threshold
scale, $T_{SUSY}^{min}$, as a function of $\sin^2\theta_W(M_Z)$
for different values of the running bottom quark
mass $m_b(M_b)$.
We also give the values of $\alpha_3(M_Z)$ and $Y_t(M_{GUT})$
which are obtained for the different set of parameters,
to show where the constraints are coming from in each case.
In particular, we observe that for the mean values
under consideration,
$\sin^2\theta_W = 0.2324$ and $m_b(M_b) = 4.3$ GeV, the
upper experimental limit on $\alpha_3(M_Z)$ is obtained
together with the upper bound on the top quark Yukawa
coupling coming from the perturbative consistency of
the theory, $Y_t(M_{GUT}) \leq 1$.

Moreover, from the numerical results, fixing the value
of $m_b(M_b)\simeq 4.3$ GeV and varying
$\sin^2\theta_W(M_Z)$, it follows that a larger (smaller)
value of $\sin^2\theta_W(M_Z)$ implies that the lower
bound on $T_{SUSY}$ comes from $\alpha_3(M_Z)$
($Y_t(M_{GUT})$).
{}From Tables 5.a-5.c, it may be noticed that, in general,
for larger values of $m_b(M_b)$ the bounds on $T_{SUSY}$ tend
to come from $\alpha_3(M_Z)$, while for smaller values of
$m_b(M_b)$ the bound tends to come from $Y_t(M_{GUT})$.
Concerning the value of the lower bound on $T_{SUSY}$, it
happens that, for any fixed value of the bottom quark mass,
a larger (smaller) value of $\sin^2 \theta_W(M_Z)$,  yields
a smaller (larger) value of $T_{SUSY}^{min}$. Indeed, such
variation
of $T_{SUSY}^{min}$ with $\sin^2\theta_W(M_Z)$ may be easily
understood. We know that $\alpha_3(M_Z)$ and $Y_t(M_{GUT})$
increase for smaller values of $T_{SUSY}$. Moreover, a larger
value of $\sin^2\theta_W(M_Z)$ implies a smaller value of
$\alpha_3(M_Z)$ and $Y_t(M_{GUT})$. Therefore, if the value
of $\sin^2\theta_W(M_Z)$ is increased then $T_{SUSY}$ has
to decrease to render $\alpha_3(M_Z)$ and $Y_t(M_{GUT})$ larger
so that one of these quantities reaches its upper bound.

It is also quite interesting to observe that, for smaller values
of the bottom quark mass, $m_b(M_b) \simeq 4.1$, which correspond
to physical masses $M_b \simeq 4.7-4.8$ GeV, the condition of
unification of couplings implies a theoretical upper bound on
$\alpha_3(M_Z)$, which turns out to be stronger than the
experimental one. In fact,
for all the experimentally acceptable
values of $\sin^2 \theta_W(M_Z)$,
if $m_b(M_b) \simeq 4.1$ GeV
perturbative unification of
gauge and Yukawa couplings may only be achieved if
$\alpha_3(M_Z) \leq 0.124$.

As we noticed in section 4, from Eq.(\ref{eq:tsusy})
we obtain that, for fixed values of the
uncolored sparticle masses, $T_{SUSY}$ decreases for
heavier colored particles.
Hence, for a fixed value of the uncolored
sparticle mass, a lower bound on $T_{SUSY}$ allows to put
an upper bound on the
values of the mass of the colored
sparticles compatible with gauge and bottom - tau Yukawa coupling
unification.
In figures 1.a-c we plot the lower bounds on
$T_{SUSY}$ in the
$m_{col}$ - $m_{wk}$ parameter space, for
fixed $m_b(M_b)$ and
different values of $\sin^2 \theta_W(M_Z)$.
Observe that, the region to the left of any of the plotted
curves is associated with
a value of $T_{SUSY}$ lower than the
allowed minimum value $T_{SUSY}^{min}$ for the given set
of parameters,  and it
is therefore ruled out. This implies that,
for a given value of the running bottom quark mass
$m_b(M_b)$ and $\sin^2\theta_W(M_Z)$,
fixing a value for the uncolored sparticle mass the
allowed values for the colored sparticle mass are
those lying to the right of the corresponding curve and the
value on the curve gives  the upper bound on $m_{col}$.
The sharp change to a constant value either for $m_{col}$
of for $m_{wk}$ corresponds to the region in which one of
these mass scales is below $M_Z$ and is hence replaced by
$M_Z$ in the expression of $T_{SUSY}$, Eq.(\ref{eq:tsusy}).
Notice that, the lower the value of $T_{SUSY}^{min}$, the less
constrained the $m_{col} - m_{wk}$  parameter space is. Thus,
from our explanation above this implies that the stronger
constraints on the values
of the colored and uncolored mass spectrum compatible with
gauge as well as bottom - tau Yukawa coupling unification
are obtained
for the smaller
values of $\sin^2\theta_W(M_Z)$ and the bottom quark mass
considered.

In section 4 we comment
on the fact that, at the two loop
level the gauge couplings receive contributions from
the Yukawa sector. In particular, the larger the
value of the top quark Yukawa coupling the larger is
its negative contribution to $\alpha_3(M_Z)$.
In fact,
for values of $Y_t(M_{GUT}) \equiv h_t^2(M_{GUT})/4\pi \simeq
0.1,$ 0.2, 0.4, 1.0, the two loop top quark Yukawa correction to
$\alpha_3(M_Z)$ is approximately
given by $\Delta^{Yuk}(\alpha_3(M_Z))
\simeq -0.001,$ -0.0015, -0.002, -0.003 respectively.
Hence,
for larger values of $Y_t(M_{GUT})$,
a slightly lower value of $T_{SUSY}$ is needed in order
to achieve unification for a given fixed
value of $\alpha_3(M_Z)$.
 The  constraints on $T_{SUSY}$ given in table 5 are based on
the condition of gauge and Yukawa coupling unification,
which, quite generally, for values of $m_b(M_b) \leq 4.6$
GeV leads to rather large values of the top quark
Yukawa coupling. If we relaxed the condition of
Yukawa coupling unification then it would be
possible to consider smaller values for $Y_t(M_{GUT})$ and,
hence, the corresponding values of $\alpha_3(M_Z)$ would
be slightly larger.
Therefore, if smaller values of $Y_t(M_{GUT})$ were taken,
the lower bound on $T_{SUSY}$ coming from the experimental
constraints on $\alpha_3(M_Z)$ would become stronger due to
the increase of the strong gauge coupling, in an amount
which can be easily worked out by
using $\Delta^{Yuk}\left(\alpha_3(M_Z)\right)$ and Eqs.
(\ref{eq:pred3})-(\ref{eq:msusy}).
For instance, the value of the strong gauge coupling
obtained for $\sin^2\theta_W(M_Z) = 0.2324$
and $T_{SUSY} = 90$ GeV
at $Y_t(M_{GUT}) \simeq 1$, which reads $\alpha_3(M_Z) \simeq
0.122$, is modified to $\alpha_3(M_Z) \simeq 0.124$
if we keep the same value of $T_{SUSY}$ and
$\sin^2\theta_W(M_Z)$ but we take a smaller value of the top
quark Yukawa coupling $Y_t(M_{GUT}) \simeq 0.1$. The latter
result is in  agreement with the value quoted in
Ref.\cite{Langacker} for the same values
of $\sin^2\theta_W(M_Z)$
and $T_{SUSY}$. The above analysis also explains the small
numerical difference observed in Tables 5.a and 5.b -
$\sin^2\theta_W(M_Z) = 0.2335$ and 0.2324 -
between the lower bounds on $T_{SUSY}$
for $m_b(M_b) = 4.3$ and 4.6 GeV, which come in both cases
from the upper experimental constraint on $\alpha_3(M_Z)$ but
involving different values of $Y_t(M_{GUT})$.

In the above discussion, we have assumed that the only splitting
in the sparticle spectrum is due to the properties of the
SUSY particles under $SU(3)_c$. However, as we mentioned before,
more complicated
situations may naturally occur.  Nevertheless,
since for a given value of $m_b(M_b)$ and $\sin^2 \theta_W$
both $Y_t(M_{GUT})$ and $\alpha_3(M_Z)$ are only dependent
on $T_{SUSY}$, our results have a more general validity,
independent of the exact supersymmetric spectrum splitting.

\section{Top Quark and Higgs Mass Predictions}

For the range of values of the running bottom quark
mass considered in the present work,
the requirement of gauge and Yukawa
coupling unification implies the convergence
of the top quark Yukawa coupling to its infrared quasi
fixed point value.
This yields accurate predictions for the running
top quark mass, which, however, depend
strongly on the range of values we take for
$\tan\beta$. In minimal supergravity models, in which
all scalar sparticles are assumed to acquire the same
soft supersymmetry breaking
mass, $m_0$, at high energy scales, constraints
on $\tan\beta$ may be obtained by requiring a proper
radiative breaking of the $SU(2)_L \times
U(1)_Y$ symmetry. In these models the characteristic
values of the ratio of the Higgs vacuum expectation values
are in the range
$1 \leq \tan\beta \leq \tan \beta^{M}$,
where $\tan \beta^M \simeq
30 (45)$ for a top quark mass $m_t \simeq 120 (180)$ GeV,
respectively.
In the following,
however, we shall be less restrictive and
allow the upper bound on $\tan\beta$ to be given
by the breakdown of the perturbative consistency of the
theory in
the bottom quark Yukawa sector, as discussed in
section 4.

  In Fig. 2 we plot the values of the running top quark
mass $m_t(m_t)$ as a function of $\tan\beta$ for different
values of $m_b(M_b)$ and taking the mean values
of $\sin^2\theta_W(M_Z) = 0.2324$
and $\alpha_3(M_Z) = 0.12$.
{}From the above figure it follows that, the smaller the value of
the bottom quark mass is, the larger the value of $m_t$ results
for any
fixed value of $\tan\beta$, in agreement with previous one loop
results \cite{KLN}. This behavior is easily understood
due to the interrelation between the top and bottom quark
Yukawa couplings. In addition, the existence of the
IR quasi fixed point for $h_t$ explains the mildness in the actual
variation of the running top quark mass values.

In Fig.3, we present the values of the running top
quark mass as a function of $\tan\beta$ for the considered mean
values of $\sin^2\theta_W(M_Z)$ and
$m_b(M_b)$  and for different values of
$\alpha_3(M_Z)$, thus, for different values of $T_{SUSY}$
and $Y_t(M_{GUT})$. For any given value of $\tan\beta$,
a larger  value of $\alpha_3(M_Z)$ corresponds to a larger
value of $m_t$. Such behavior is predictable due to the
interrelation between the strong gauge coupling constant and
the top quark Yukawa coupling. However,
it is worth noticing
that, although, for example, for
$\tan\beta = 10$, a variation in the low energy value of
the strong coupling constant, $\alpha_3(M_Z) = 0.115
\rightarrow 0.130$ induces a large variation in $Y_t(M_{GUT})
= 0.28 \rightarrow 1$,  due to the presence of the
infrared quasi fixed point this
translates in an increase in $m_t$ of
about a five percent of its original value.

Quite generally, for values of $\tan \beta \leq 10$, the top
quark mass is a monotonically increasing function of
$\tan \beta$, varying from $m_t = 0$ for $\tan \beta =0$ to
a  value of $m_t$ far
above its lower experimental bound, for
$\tan \beta \simeq 10$. Hence, the experimental lower bound
 on the top quark mass,
allows to put a lower bound on $\tan \beta$, below which
the model is inconsistent with experimental data.
As a matter of fact, for
the range of values of $\alpha_3(M_Z)$ and $m_b(M_b)$
considered in this work, it follows that
$\tan \beta \geq 0.4$. Observe
that a top quark mass $M_t \simeq 138$ GeV, which provides
the best fit for the experimental data, may only be achieved
either
for values of $\tan\beta$  close to one, or for very large
values of $\tan\beta$,
close to the upper bound on this quantity.

 A phenomenologically interesting possibility is the unification
of the  three
Yukawa couplings of the third generation.
Since the RG equations
of the top and bottom quark Yukawa couplings differ only in
electroweak terms, the unification of these two
couplings may only be achieved for values of $\tan\beta
\simeq m_t(m_t)/m_b(m_t)$. From the behavior of $m_t(m_t)$
as a function of $\tan\beta$ and the fact that $m_b(m_t)
\simeq 3$ GeV, it is easy to see that the unification of
the three Yukawa couplings occurs for values of
 $\tan \beta \simeq
60$. For example, for $\sin^2 \theta_W(M_Z)
\simeq 0.2324$, $\alpha_3(M_Z) \simeq  0.122$ and
$m_b(M_b) = 4.1,4.3,4.6$ GeV, the unification of the
tau, bottom and top Yukawa
couplings occurs for $\tan\beta \simeq 64,60,53$, respectively
(see tables 1.a - 1.c).

Concerning the Higgs sector,
as we mentioned in section 5, the light
Higgs mass predictions
are not only dependent  on the value of the top quark
Yukawa coupling,
but on the absolute value of the squark mass as well.
In fact,
for $m_{\tilde{q}} \gg m_t$, the dominant
radiative corrections to the tree level Higgs mass
value, $m_h^2 =
M_Z^2 \cos 2 \beta$, are proportional to the logarithm of
the ratio of the top and stop masses, with a factor which
depends on the fourth power of the top quark
Yukawa coupling \cite{OYY}-\cite{HH},
\begin{equation}
m_h^2(m_t) \simeq M_Z^2 \cos^2 2 \beta + \frac{3}{2\pi^2}
v^2 \sin^4\beta \;
h_t^4(m_t) \; \ln \left(\frac{m_{\tilde{q}}}{m_t}
\right) .
\end{equation}
Since the unification condition only fixes the value of
the effective supersymmetric threshold scale $T_{SUSY}$,
which is only weakly dependent on the squark masses,
even if we fix the
value of $\tan\beta$, the lightest
CP even Higgs mass can not be predicted within this
scheme. However, an upper bound on $m_{\tilde{q}}$ may
always be obtained by the additional requirement of the
stability of the hierarchy of scales under radiative
corrections, or equivalently, the absence of fine
tuning of the soft supersymmetry breaking parameters in
order to achieve the proper weak scale. This upper bound
is, quite generally, of the order of a few TeV. Hence,
even though the Higgs mass scale may not be directly
determined from the unification condition, an upper
bound on it may be obtained,
which, for a given $T_{SUSY}$ is achieved for the maximum
allowed value of $m_{\tilde{q}}$.

In Figure 4, we plot the Higgs mass as a
function of $\tan\beta$ for $\sin^2 \theta_W(M_Z) =
0.2324$, $m_b(M_b) \simeq 4.3$ GeV,
$T_{SUSY} \simeq 1 TeV$, thus $\alpha_3(M_Z)
\simeq 0.115$,
and for  different values of the colored and
uncolored particles.  Observe that, as expected, the
Higgs mass increases with the colored sparticle mass.
For $m_{\tilde{q}} = 1,3,10$ TeV,
the  lightest CP even Higgs mass, as a function of $\tan\beta$
reads $m_h = 54,67,78$ GeV for
$\tan \beta = 1$; $m_h = 111,128,143$ GeV for
$\tan\beta = 3$ and $m_h = 126,142,158$ GeV for
$\tan\beta = 10$, respectively.

\section{Conclusions}

We have analyzed in detail the conditions under
which gauge and Yukawa coupling
unification take place within
the MSSM. We derived an analytical expression for the
effective supersymmetric threshold scale as a function
of the supersymmetric particle masses. We obtain that the
effective  scale $T_{SUSY}$
is only slightly dependent on the squark and slepton
masses, while it is strongly dependent on the Higgsino
and gaugino masses. In particular, we have shown that,
in a first approximation - whenever  no large
mixing between the Higgsinos and the weak
gauginos occurs,  the
gaugino masses proceed from a common soft supersymmetry
breaking term at $M_{GUT}$, and
all sparticle masses are above $M_Z$ - the supersymmetric
threshold scale is given by the overall Higgsino mass
multiplied by a factor which depends on the ratio of
the strong and the weak coupling constants. This result
is quite interesting,  due to the fact that,
if there is no significant mixing, the Higgsino mass is
governed by the supersymmetric mass parameter $\mu$
appearing in the superpotential and, hence, there
is no explicit dependence on the soft supersymmetry breaking
terms in the expression of $T_{SUSY}$.
Quite generally, our analysis is made under the assumption
that all SUSY particles have masses above $M_Z$. We have,
however, analyzed the modification of the expression for
the effective supersymmetric threshold scale
in the case in which  light sparticles appear
in the spectrum as well.

We show  that, not only the value of the strong gauge coupling,
but also the top quark Yukawa sector depends dominantly on
$T_{SUSY}$. For a given set of values for $\sin^2\theta_W$
and the bottom quark mass, we derive lower bounds on $T_{SUSY}$
coming either from the experimental upper bounds on
 $\alpha_3(M_Z)$ or from the requirement of perturbative
consistency of the top Yukawa sector of the theory.
Furthermore, within a
simplified scheme, in which we assign a common mass $m_{wk}$ to
the uncolored sparticles and a mass $m_{col}$ to the colored
ones, a lower bound on $T_{SUSY}$ determines an upper bound on
the value of  $m_{col}$, for
any given value of $m_{wk}$.

Performing a two loop renormalization group analysis,
we have also computed  the predictions for the top quark
and Higgs masses as a function of $\tan\beta$ in the
framework of gauge and bottom-tau Yukawa coupling unification.
Quite generally,  for different
values of $\sin^2\theta_W (M_Z)$ and $m_b(M_b)$ and
for various values of $\alpha_3(M_Z)$, we obtain that
the running top quark mass has to be below 200 GeV, while
a value of
$m_t \simeq 140$ GeV is only compatible with values of
$\tan\beta$ quite close to one, or with extremely large values
of $\tan\beta \simeq 60$.  Contrary to what happens with
$\alpha_3(M_Z)$ and $m_t$, for fixed values of the bottom quark
mass and the weak mixing angle, the light
Higgs mass depends not only
on $T_{SUSY}$, but on the value of the top squark mass as well.
For $m_{\tilde{q}} \simeq 1$ TeV, for example, the Higgs mass
varies from values which are close to its experimental lower
bound if $\tan\beta \simeq 1$ to values of $m_h \simeq 130$
GeV for $\tan\beta \simeq 10$.

Finally, we should mention that
in the present work, we have not considered the effects of
possible threshold corrections to the gauge and Yukawa couplings
arising at the grand unification scale. These corrections can
only be computed once we know the specific grand unified scenario
at the high energy scales. In spite of that, under general
assumptions,
a somewhat general parametrization of
such corrections may be done. In Ref.\cite{Langacker}, for
example, the high energy scale threshold
corrections associated with the gauge coupling
constants were considered. We should stress that, within the
framework of gauge and Yukawa coupling unification, not only
the threshold corrections to the gauge couplings, but also those
to the Yukawa couplings must be considered. In fact,
if the Yukawa threshold corrections were as large as
5 to 10$\%$, they would have relevant
effects on the top quark
mass predictions.   We shall concentrate on this subject  in a
forthcoming publication \cite{Prep}. \\
{}~\\
ACKNOWLEDGEMENTS\\
{}~\\
M.C. and C.W. would like to thank W. A. Bardeen  and P. Ramond
for useful conversations, which motivated part of this
study, and L. Clavelli and H. P. Nilles for interesting comments.
S. P. is  supported in part by the
Polish Committee for Scientific Research.

\newpage
\hspace*{-0.6cm} {\bf Table  Captions.}  \\
{}~\\
Table 1.a. Comparison between the predictions obtained within
the two loop RG analysis, in the framework of
gauge and Yukawa coupling unification, and the approximate
results derived from the infrared quasi fixed point predictions,
for $\sin^2 \theta_W(M_Z) = 0.2324$,
$\alpha_3(M_Z) \simeq 0.122$ and  $m_b(M_b) = 4.6$ GeV.\\
{}~\\
Table 1.b. The same as in Table 1.a but for
$m_b(M_b) = 4.3$ GeV.  \\
{}~\\
Table 1.c. The same as in Table 1.a but for
 $m_b(M_b) = 4.1$ GeV. \\
{}~\\
Table 2. Dependence of $\alpha_3(M_Z)$ on $\sin^2 \theta_W(M_Z)$
and on the effective supersymmetric threshold scale $T_{SUSY}$
in the framework of the two loop RG analysis with gauge and
Yukawa coupling unification,
for $m_b(M_b) = 4.3$ GeV.\\
{}~\\
Table 3. Dependence of $Y_t(M_{GUT})$ on $\alpha_3(M_Z)$
and $T_{SUSY}$, in the framework of gauge and
Yukawa coupling unification. The results are for $\sin^2
\theta_W(M_Z) = 0.2324$ and $m_b(M_b) = 4.3$ GeV, while fixing
$m_{col} = 1$ TeV and varying $m_{wk}$ to get the corresponding
value of the supersymmetric threshold scale.\\
{}~\\
Table 4. Top quark Yukawa coupling and Higgs mass
dependence on the precise values of
$m_{col}$ and $m_{wk}$ for a
fixed value of $T_{SUSY} = 1$ TeV.
The results are for $\sin^2\theta_W(M_Z) = 0.2324$,
$m_b(M_b) = 4.3$ GeV and fixing
$\tan\beta = 4$. \\
{}~\\
Table 5.a. Lower bounds on the effective supersymmetric threshold
scale as a function of the running bottom quark mass for:
$\sin^2\theta_W(M_Z) = 0.2335$.
The values of $\alpha_3(M_Z)$
and $Y_t(M_{GUT})$ obtained for each particular set of
low energy parameters are also given.\\
{}~\\
Table 5.b. The same as in Table 5.a. but for $\sin^2\theta_W(M_Z)
= 0.2324$.\\
{}~\\
Table 5.c. The same as in Table 5.a. but for $\sin^2\theta_W(M_Z)
= 0.2315$.\\

\newpage
\hspace*{-0.6cm} {\bf Figure Captions.}\\
{}~\\
Fig. 1.a.  Lower bound on $T_{SUSY}$ as a function of the
colored and uncolored sparticle mass scales, $m_{col}$ and
$m_{wk}$, respectively, for a running bottom quark mass
$m_b(M_b) = 4.6$ GeV and for
$\sin^2\theta_W(M_Z) = 0.2315$ (dashed line),
$\sin^2\theta_W(M_Z) = 0.2324$ (dot-dashed line)
and  $\sin^2\theta_W(M_Z) = 0.2335$ (solid line).\\
{}~\\
Fig. 1.b. The same as figure 1.a. but for $m_b(M_b) = 4.3$ GeV.\\
{}~\\
Fig. 1.c. The same as figure 1.a. but for $m_b(M_b) = 4.1$ GeV.\\
{}~\\
Fig. 2. The running top quark mass as a function of $\tan\beta$
for $\sin^2\theta_W(M_Z) = 0.2324$, $\alpha_3(M_Z) \simeq 0.122$
and varying the
bottom
quark mass to be $m_b(M_b) = 4.1$ GeV (dot-dashed line),
$m_b(M_b) = 4.3$ GeV (solid line) and
$m_b(M_b) = 4.6$ GeV (dashed line). \\
{}~\\
Fig. 3. The running top quark mass as a function of $\tan\beta$
for $m_b(M_b) = 4.3$ GeV and $\sin^2\theta_W(M_Z) = 0.2324$ and
varying the value of the strong gauge coupling constant to be
$\alpha_3(M_Z) = 0.115$ (solid line), 0.122 (dashed line) and
0.13 (dot-dashed line). \\
{}~\\
Fig. 4. The light Higgs mass as a function of $\tan\beta$ for
$\sin^2\theta_W = 0.2324$ and $m_b(M_b) = 4.3$ GeV while varying
the value of the colored sparticle mass scale to be
$m_{col} = 1$ TeV
(dashed line), 3 TeV (dot-dashed line) and 10 TeV (solid line)
and taking $m_{wk}$ accordingly to fix $T_{SUSY} = 1$ TeV.

\newpage
\begin{center}
\begin{tabular}{|c|c|c|}
\hline \
$\;\;\;\;\;\;\;\;\;\;\;\;\;$
&IR Fixed Point &RG solution
\\  \hline
$m_t(\tan\beta=1)
$ [GeV]
 &144  &141
\\ \hline
$m_t(\tan\beta=3)$ [GeV]
 &192  &184
\\ \hline
$m_t(\tan\beta=10)$ [GeV]
 &201  &192
\\ \hline
$m_t(h_t=h_b)$ [GeV]
 &188  &170
\\ \hline
$\tan \beta
(h_t=h_b)$
 &58 &53
\\ \hline
\end{tabular}
\\
{}~\\
Table 1.a
\end{center}
{}~\\

\begin{center}
\begin{tabular}{|c|c|c|}
\hline \
$\;\;\;\;\;\;\;\;\;\;\;\;\;$
&IR Fixed Point &RG solution
\\  \hline
$m_t(\tan\beta=1)
$ [GeV]
 &144  &144
\\ \hline
$m_t(\tan\beta=3)
$ [GeV]
 &192  &187
\\ \hline
$m_t(\tan\beta=10)$ [GeV]
 &201  &196
\\ \hline
$m_t(h_t=h_b)$ [GeV]
 &188  &180
\\ \hline
$\tan \beta
(h_t=h_b)$
 &63   &60
\\ \hline
\end{tabular}
\\
{}~\\
Table 1.b
\end{center}
{}~\\

\begin{center}
\begin{tabular}{|c|c|c|}
\hline \
$\;\;\;\;\;\;\;\;\;\;\;\;\;$
&IR Fixed Point &RG solution
\\  \hline
$m_t(\tan\beta=1)
$ [GeV]
 &144  &144
\\ \hline
$m_t(\tan\beta=3)$ [GeV]
 &192  &188
\\ \hline
$m_t(\tan\beta=10)$ [GeV]
 &201  &197
\\ \hline
$m_t(h_t=h_b)$ [GeV]
 &188  &185
\\ \hline
$\tan \beta
(h_t=h_b)$
 &66 &64
\\ \hline
\end{tabular}
\\
{}~\\
Table 1.c
\end{center}
{}~\\

\begin{center}
\begin{tabular}{|c|c|c|}
\hline \
$\sin^2 \theta_W(M_Z)$
&$\alpha_3(M_Z)$ for $T_{SUSY} = 1$ TeV
 &$\alpha_3(M_Z)$ for $T_{SUSY} = 100$ GeV
\\  \hline
$0.2335
$
 &0.111  &0.118
\\ \hline
$0.2324$
 &0.115  &0.122
\\ \hline
$0.2315$
 &0.118  &0.126
\\ \hline
\end{tabular}
\\
{}~\\
Table 2.
\end{center}
{}~\\

\begin{center}
\begin{tabular}{|c|c|c|c|}
\hline \
$m_{wk}$[GeV]
&$T_{SUSY}$[GeV] &$\alpha_3(M_Z)$   &$Y_t(M_{GUT})$
\\  \hline
$10^3$
 &$10^3$  &0.115   &0.2
\\ \hline
200
 &15   &0.127   &0.7
\\ \hline
$150$
 &7    &0.130   &1.0
\\ \hline
\end{tabular}
\\
{}~\\
Table 3.
\end{center}
{}~\\

\begin{center}
\begin{tabular}{|c|c|c|c|c|c|}
\hline \
$m_{col}$ [TeV]
&$m_{wk}$[TeV]    &$Y_t(M_{GUT})$
&$\alpha_3(M_Z)$   &$m_t$ [GeV]    &$m_h$ [GeV]
\\  \hline
1
 &1  &0.28    &0.115  &188  &118
\\ \hline
3
 &1.93 &0.29    &0.115 &191  &135
\\ \hline
10
 &3.89 &0.33   &0.115 &193 &150
\\ \hline
\end{tabular}
\\
{}~\\
Table 4.
\end{center}
{}~\\

\begin{center}
\begin{tabular}{|c|c|c|c|}
\hline \
$sin^2 \theta_W(M_Z) =0.2335$
&$T^{min}_{SUSY}$[GeV] &$\alpha_3(M_Z)$   &$Y_t(M_{GUT})$
\\  \hline
$m_b(M_b) = 4.6$ GeV
 &2.5  &0.13    &0.3
\\ \hline
$m_b(M_b) = 4.3$ GeV
 &1.5  &0.13    &0.9
\\ \hline
$m_b(M_b) = 4.1$ GeV
 &12   &0.123   &1.0
\\ \hline
\end{tabular}
\\
{}~\\
Table 5.a
\end{center}
{}~\\

\begin{center}
\begin{tabular}{|c|c|c|c|}
\hline \
$\sin^2 \theta_W(M_Z) = 0.2324$
&$T^{min}_{SUSY}$[GeV] &$\alpha_3(M_Z)$   &$Y_t(M_{GUT})$
\\  \hline
$m_b(M_b) = 4.6$ GeV
 &9    &0.13    &0.34
\\ \hline
$m_b(M_b) = 4.3$ GeV
 &6.5  &0.13   &1.0
\\ \hline
$m_b(M_b) = 4.1$ GeV
 &90   &0.122   &1.0
\\ \hline
\end{tabular}
\\
{}~\\
Table 5.b
\end{center}
{}~\\

\begin{center}
\begin{tabular}{|c|c|c|c|}
\hline \
$\sin^2 \theta_W(M_Z)=0.2315$
&$T^{min}_{SUSY}$[GeV] &$\alpha_3(M_Z)$   &$Y_t(M_{GUT})$
\\  \hline
$m_b(M_b) = 4.6$ GeV
 &25   &0.13    &0.4
\\ \hline
$m_b(M_b) = 4.3$ GeV
 &45   &0.128  &1.0
\\ \hline
$m_b(M_b) = 4.1$ GeV
 &350  &0.121   &1.0
\\ \hline
\end{tabular}
\\
{}~\\
Table 5.c
\end{center}

\newpage
\begin{thebibliography}{99}
\bibitem{earlySM} H. Georgi and S. L. Glahsow, Phys. Rev.
Lett. 32 (1974) 438.
\bibitem{early} H. Georgi, H. Quinn and S. Weinberg, Phys.
Rev. Lett. 33 (1974) 451.\\
S. Dimopoulos and H. Georgi, Nucl. Phys. B193 (1981) 150.\\
N. Sakai, Z. Phys. C11, 153 (1981)\\
E. Witten, Nucl. Phys. B188, 513 (1981).
\bibitem{ABF} J. Ellis, S. Kelley and D.V. Nanopoulos,
Phys. Lett. B260 (1991) 131;\\
P. Langacker and M.X. Luo, Phys. Rev. D44 (1991) 817;\\
F. Anselmo, L. Cifarelli, A. Peterman and
A. Zichichi, Nuovo Cimento 104A (1991) 1817;\\
U. Amaldi, W. de Boer and H. F\'urstenau, Phys.
Lett. B260 (1991) 447.
\bibitem{BB} W. A. Bardeen, A. J. Buras, D. W. Duke and
T. Muta, Phys. Rev. D 18 (1978) 3998.
\bibitem{EKN} J. Ellis, S. Kelley and D. V. Nanopoulos,
Phys. Lett. B249 (1990) 441; Phys. Lett. B287 (1992) 95;
Nucl. Phys. B373 (1992) 55;\\
F. Anselmo, L. Cifarelli, A. Peterman and A. Zichichi,
Nuovo Cimento 105A (1992) 581;\\
R. Barbieri and L. Hall, Phys. Rev. Lett. 68 (1992) 752;\\
J. Hisano, H. Murayama and T. Yanagida, Phys. Rev. Lett.
69 (1992) 1014.
\bibitem{RR} G. G. Ross and R. G. Roberts, Nucl. Phys. B377
(1992) 571.
\bibitem{Langacker} P. Langacker and N. Polonsky, U. of
Pennsylvania preprint, UPR-0513T, October 1992.
\bibitem{AN} R. Arnowitt and P. Nath, Phys. Rev. Lett.
69 (1992) 725; P. Nath and R. Arnowitt, Phys. Lett. B287
(1992) 89; ibid B289 (1992) 368.
\bibitem{LNPZ} J. Lopez, D.V. Nanopoulos, H. Pois and
A. Zichichi, Phys. Lett. B299 (1993) 262.
\bibitem{Ramond} H. Arason, D. J. Casta\~no, B. Keszthelyi,
S. Mikaelian, E. J. Piard, P. Ramond and B. D. Wright,
Phys. Rev. Lett. 67 (1991), 2933.
\bibitem{KLN} S. Kelley, J.L. Lopez and D.V. Nanopoulos,
Phys. Lett. B278 (1992) 140.
\bibitem{DHR} S. Dimopoulos, L. Hall and S. Raby,
Phys. Rev. Lett. 68 (1992) 1984, Phys. Rev. D45 (1992) 4192.
\bibitem{DFS} G. Degrassi, S. Fanchiotti and A. Sirlin,
Nucl. Phys. B351 (1991) 49.
\bibitem{Glu} L. Clavelli, Phys. Rev. D46 (1992) 2112 \\
M. Jezabek and J. K\"uhn, Karlsruhe U. preprint, TTP-92-37.
\bibitem{Altarelli} G. Altarelli, CERN preprint, CERN-TH-6623-92.
\bibitem{Partd} K. Hikasa et al, Particle Data Group,
Phys. Rev. D 45 (1992).
\bibitem{RSM} H. Arason,
D. J. Casta\~no, B. Keszthelyi,
S. Mikaelian, E. J. Piard, P. Ramond and B. D. Wright,
Phys. Rev. D 46 (1992) 3945.
\bibitem{SCJN} W. Siegel, Phys. Lett. B84 (1979) 193;\\
D.M. Capper, D.R.T. Jones and P. van Nieuwenhuizen, Nucl.
Phys. B167 (1980) 479.
\bibitem{Prep} M. Carena, S. Pokorski and C. Wagner, in preparation.
\bibitem{KLNQ} See, for example, C. Kounnas, A.B. Lahanas,
D.V. Nanopoulos and M. Quiros, Nucl. Phys. B236 (1984) 438.
\bibitem{OYY} Y. Okada, M. Yamaguchi and T. Yanagida, Prog. Theor.
Phys. 85 (1991) 1; Phys. Lett. B262 (1991) 54.
\bibitem{ERZ} J. Ellis, G. Ridolfi and F. Zwirner, Phys.
Lett. B257 (1991) 83.
\bibitem{HH} H. Haber and R. Hempfling, Phys. Rev. Lett.
66(1991) 83.
\bibitem{BF} R. Barbieri, M. Frigeni and F. Caravaglios,
Phys. Lett. B258 (1991) 167.\\
R. Barbieri and M. Frigeni, Phys. Lett. B258 (1991) 395.
\bibitem{ST} P. Chankowski, S. Pokorski and J. Rosiek,
Phys. Lett. B274 (1992) 191.
\bibitem{Dyn} M. Carena, T.E. Clark, C.E.M. Wagner, W.A. Bardeen
and K. Sasaki, Nucl. Phys. B369 (1992) 33.
\bibitem{Higgs} M. Carena, K. Sasaki and C.E.M. Wagner, Nucl.
Phys. B381 (1992) 66.
\bibitem{Stefan} P. Chankowski, S. Pokorski and J. Rosiek,
Phys. Lett. B281 (1992) 100.
\bibitem{HH2} H. Haber and R. Hempfling, UC, Santa Cruz
preprint, SCIPP-92-33, submitted to Phys. Rev. D.
\bibitem{CH} P. Chankowski, Phys. Rev. D41 (1990) 2877.
\bibitem{RG} D. R. T. Jones, Phys. Rev. D25 (1982) 581.\\
J.E. Bj\"orkman and D.R.T. Jones, Nucl. Phys. B259 (1985) 533.
\bibitem{RGSM} M. E. Machacek and M. T. Vaughn, Nucl. Phys.
B222 (1983) 83; Nucl. Phys. B236 (1984) 221; Nucl. Phys.
B249 (1985) 70.
\bibitem{IR} C. T.
Hill, Phys. Rev. D24 (1981) 691;\\
C. T. Hill, C. N. Leung and S. Rao, Nucl. Phys.
B262 (1985) 517.


\end{thebibliography}
\end{document}